\documentclass{article}

\usepackage{arxiv}

\usepackage{booktabs}
\usepackage[utf8]{inputenc} 
\usepackage[T1]{fontenc}    
\usepackage{hyperref}       
\usepackage{url}            
\usepackage{booktabs}       
\usepackage{amsfonts, amsmath, amssymb, amsthm, bbm}
\usepackage{nicefrac}       
\usepackage{microtype}      
\usepackage{lipsum}
\usepackage{fancyhdr}       
\usepackage{graphicx}       
\graphicspath{{media/}}     
\usepackage[export]{adjustbox}
\usepackage{todonotes}
\usepackage{tikz}
\usetikzlibrary{patterns}
\usepackage{algorithm} 
\usepackage[noend]{algpseudocode}
\def\code#1{\texttt{#1}}
\newcommand{\pluseq}{\mathrel{+}=}
\newcommand{\minuseq}{\mathrel{-}=}
\DeclareMathOperator*{\argmax}{arg\,max}

\title{A Modular Framework\\ for Reinforcement Learning Optimal Execution}

\author{ 
	\href{https://orcid.org/0000-0003-2233-1858}{\includegraphics[scale=0.06]{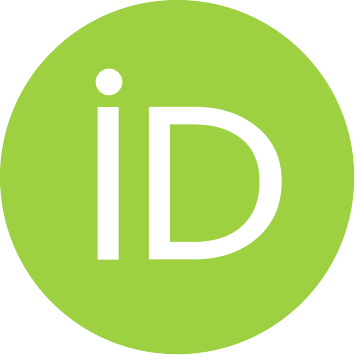}\hspace{1mm}Fernando de Meer Pardo \thanks{\textit{corresponding author}}}\\
	School of Engineering\\
	Zurich University of Applied Sciences\\
	Switzerland\\
	\texttt{demp@zhaw.ch}\\
	\And 
    \href{https://orcid.org/0000-0003-0824-5267}{\includegraphics[scale=0.06]{orcid.pdf}\hspace{1mm}Cristoph Auth}\\
	School of Engineering\\
	Zurich University of Applied Sciences\\
	Switzerland \\
	\texttt{christoph.auth@warwickgrad.net} \\
	\And
	\href{https://orcid.org/0000-0002-9227-2438}{\includegraphics[scale=0.06]{orcid.pdf}\hspace{1mm}Florin Dascalu} \\
	School of Management and Law\\
	Zurich University of Applied Sciences\\
	Switzerland\\
	\texttt{dasc@zhaw.ch} \\
}

\date{}

\renewcommand{\headeright}{}
\renewcommand{\undertitle}{}
\renewcommand{\shorttitle}{A Modular Framework for RL Optimal Execution}

\hypersetup{
pdftitle={A Modular Framework for Reinforcement Learning Optimal Execution},
pdfsubject={q-bio.NC, q-bio.QM},
pdfauthor={Fernando De Meer Pardo, Christoph Auth, Florin Dascalu},
pdfkeywords={First keyword, Second keyword, More},}

\begin{document}
\maketitle

\begin{abstract}
	In this article, we develop a modular framework for the application of Reinforcement Learning to the problem of Optimal Trade Execution. The framework is designed with flexibility in mind, in order to ease the implementation of different simulation setups. Rather than focusing on agents and optimization methods, we focus on the environment and break down the necessary requirements to simulate an Optimal Trade Execution under a Reinforcement Learning framework such as data pre-processing, construction of observations, action processing, child order execution, simulation of benchmarks, reward calculations etc. 
	We give examples of each component, explore the difficulties their individual implementations \& the interactions between them entail, and discuss the different phenomena that each component induces in the simulation, highlighting the divergences between the simulation and the behavior of a real market. We showcase our modular implementation through a setup that, following a Time-Weighted Average Price (TWAP) order submission schedule, allows the agent to exclusively place limit orders, simulates their execution via iterating over snapshots of the Limit Order Book (LOB), and calculates rewards as the \$ improvement over the price achieved by a TWAP benchmark algorithm following the same schedule. We also develop evaluation procedures that incorporate iterative re-training and evaluation of a given agent over intervals of a training horizon, mimicking how an agent may behave when being continuously retrained as new market data becomes available and emulating the monitoring practices that algorithm providers are bound to perform under current regulatory frameworks.
\end{abstract}


\section{Introduction}

An execution in an electronic financial market context is the process in which a number of units of a publicly traded financial asset are bought or sold through said market. Optimal Execution algorithms aim to carry out this process by dividing the transaction into child orders in such a way that a series of metrics, such as the transaction costs incurred, the order slippage (the effect of order submissions on market prices) and/or the \textit{Implementation Shortfall} (IS) (the difference between market prices at the start and end of the transaction, referred to this way because it represents the implementation cost difference between ideal and realized portfolios) are optimized.\\

Modern electronic financial markets are based on trading pools that implement a continuous double auction mechanism. Market participants are able to send two types of orders. Limit orders represent an open interest to buy/sell at a maximum/minimum price. If they fail to instantly lead to a transaction, they are entered into a market structure called the Limit Order Book (LOB) that stores all limit orders and that matches compatible incoming orders of the opposite side based on a “first in, first out” queuing system specific to each price-level \cite{MarketBeahviourClearingHouse1982,MarketMakingbook1985, DOMOWITZ199429,HFSimulationsofLOBs2011}. Market orders, on the contrary, represent an interest to buy or sell at the best available price determined by the limit orders currently in the LOB. Limit orders guarantee a price but not the execution and  provide the market with liquidity, allowing other market participants the chance to trade at known prices. In contrast, market orders guarantee the execution but not the price at which it is resolved therefore removing liquidity from the market.\\

Since the profitability of any investment heavily depends on the effectiveness of its trade execution, in some jurisdictions explicit rulings regulate the obligation of financial service providers to perform optimal/best execution for their clients. The main aim of these rulings is to prevent financial service providers from collecting rebates on transactions by executing client orders at more unfavorable prices that those present in the public markets, but they also force providers to choose a method of execution if the client has not explicitly done so. Examples of this are the US Financial Industry Regulatory Authority (FINRA) Rule $5310$ \cite{USFINRA5310} which states that abiding members shall "buy or sell [...] so that the resultant price to the customer is as favorable as possible under prevailing market conditions" and the  EU's Markets in Financial Instruments Directive (MiFID) Article 21 of the L1 Directive \cite{eu-MiFID} which states that "firms should monitor on a regular basis the effectiveness of their [\textit{best execution}] policy and correct any deficiencies identified".\\

In other jurisdictions, regulatory bodies issue guidance laying out requirements and expected good practices regarding best execution, as well as showcase common practices that fail to meet standards. Examples of this are the Swiss Financial Markets Authority (FINMA) Circular 2018/1 \cite{FINMACircular2018}, Hong Kong's Securities and Futures Commission (SFC) 2018 Best Execution Report \cite{SFCReport2018}, and the Monetary Authority of Singapore's (MAS) 2020 Notice on Execution of Client Orders \cite{MASNoticeonExecution2020}. All of these directives encourage financial service providers to put in place operational mechanisms to continuously monitor execution quality, document phenomena such as execution delays, disclose specific execution arrangements to clients, and designate specific committees in which both management, regulatory compliance units, and the teams responsible of carrying out client executions collaborate on policy oversight.\\

This set of obligations, explainability concerns, early results in the study of the Optimal Execution problem (see Bertsimas \& Lo \cite{Bertsimas1998OptimalCO} and Almgren \& Chriss \cite{almgren2000optimal}), and specially the fact that the monitoring of performed executions has to be carried out via comparison against market benchmarks such as the Volume Weighted Average Price (VWAP) are among the factors that have contributed to the domination of a small number of well-known algorithms over the commercial offering of execution services \cite{NasDAQNordicAlgoOffer}. Examples include algorithms that attempt to directly capture the VWAP benchmark such as Percentage-Of-Volume algorithms that aim to participate in a given $\%$ of market volume \cite{Kakade2004CompetitiveAF}, Time Weighted Average Price (TWAP) algorithms that submit child orders of the same volume at a constant rate, and algorithms that interact with the bid/ask spread \cite{EffectiveTradeExe2012Cesari}. All of these algorithms incorporate a relatively limited amount of market information in their decision-making process and decide order submissions based on relatively straightforward easily explainable rules.

\subsection{Related Work}\label{Related Work}

 Bertsimas \& Lo \cite{Bertsimas1998OptimalCO} and Almgren \& Chriss \cite{almgren2000optimal} are perhaps the most influential seminal works on the mathematical treatment of Optimal Execution. The authors conceive the market price process as a realization of an arithmetic random walk and model market impacts (both temporary and permanent) via a family of functions whose parameters can be estimated with historical data. They obtain the optimal strategy, defined as the strategy that minimizes the expected cost of trading under their assumptions (and considering only market orders), via Dynamic Programming and Stochastic Control respectively. Extensions to the original approach that model additional market phenomena via stochastic processes include \cite{Almgren2003OptimalEW,Almgren2008OptimalTI, Dang2014OptimalEW, FORSYTH20121971, FORSYTH2011241,RiskAversionDyn2009, cartea2017a}. Another branch of the literature focuses instead on the modelling of the order book directly  \cite{OptPortLiqLimitOrders2011, 2011LiqLimitOrderBooks2011, OBIZHAEVA20131, OptExeStratLOBsGeneralShapeFunctions2007, OptExegeneral1SidedLOB2011, OptLiquidityTrading2005}, again via stochastic modelling and solved via stochastic control techniques. The main drawback of the stochastic modelling paradigm is \textit{model risk}, i.e., the choice of model depends on stringent assumptions, which may not hold in reality, on the nature of the market phenomena being modelled. For example, assumptions such as the independence of returns and linear market impact functions lead to the naïve TWAP (placing equally sized orders over the execution) being the only optimal strategy, which is difficult to justify across all markets/assets in a real setting. Furthermore, the chosen functional forms of the stochastic processes used in models often lead to unavailable analytic solutions that only allow for approximations, costly parameter calibrations, and/or complex optimization problems which can become computationally intractable and/or run into convergence challenges.\\
 
 \cite{RLForOptimalTradeExe2006} was the first work to explore the application of data-driven Reinforcement Learning by recursively employing Q-learning and dynamic programming on High Frequency market data. Later, general advances in Deep Learning led to a new era of Reinforcement Learning research in which Neural Networks started being used as agents \cite{PlayingAtari2013DeepMind, OpenAIGymPaper2016}, and hence emerged a new Optimal Execution paradigm based on Deep Reinforcement Learning (DRL). Contrary to stochastic modelling, Deep Reinforcement Learning allows for a data-driven, model-free approach to Optimal Execution in which we obtain strategies through learning optimal policies for sequential decision making by optimizing a cumulative reward function. The only necessary assumption is the Markov property of the system at hand, i.e., that the optimal action at any given state of the environment is independent of previous actions, which is appropriate for our setting since we can include the results of previous actions as internal variables of the state. A multitude of different setups have been explored in the DRL literature, including Q-learning \cite{DoubleDeepQLearn2018, lin2019optimal,SCHNAUBELT2022993, DeepExeBeatingMarketBenchmarks2019, karpe_multi-agent_2020,ARLExtensiontotheACFramework2014}, Policy Gradient Methods \cite{2019MultiAgentRLLiqStrat, EndtoEndPaper2020, PolicyDistillation2021, SCHNAUBELT2022993, OptTradeExeRL2018, DeepExeBeatingMarketBenchmarks2019}, and $\text{Sarsa}(\lambda)$ \cite{OptTradeExeRL2018}, among others.\\
 
The advantages of the Deep RL paradigm include being able to obtain, by learning from experience on historical market data, adaptable execution policies that are less susceptible to front-running and avoid some aspects of model risk. Model choices in DRL are also inevitable, but they are usually not made based on assumptions about the nature of the underlying system to optimize, but rather on the empirical performance of the combination of the Neural Network Architecture used as agent and the RL optimization algorithm. The main disadvantages include the black-box nature of DRL execution strategies and the difficulty that financial simulations, that constitute the training environment for policies, involve.\\

Most of the works in the literature focus on the Neural Network Architectures used as agents and the RL optimization methods employed, as it is common in other areas of RL research. However, one crucial characteristic of RL applied in market applications is that the true environment is inaccessible and thus it is necessary to simulate its behaviour, be it via historical data or data generation approaches. This challenge constitutes an additional level of choice compared to RL settings with access to their true environment. Different setups differ not only by the NN agents and RL optimization methods they employ, but also by the type of market simulation they implement (that is, by the environment of their RL setup). The performance of a given agent + optimization method does not necessarily generalize to different simulation setups and the ability of a setup to replicate the behavior of a real market is in itself a major challenge. The works mentioned above differ not only by the RL optimization methods mentioned to categorize them, but also by the multitude of choices they make in order to simulate executions.\\

\subsection{Our Contributions}

We focus on the environment/simulation setup aspect of RL Optimal Execution. Few works openly share the implementation of their proposed setup along with the training data used, and often the hyper-parameters employed during the experiments are not fully specified, leading to the impossibility of implementing the simulation setups and reproducing results. Reproducible setups are routinely benchmarked against proposed novel setups but often implementation choices are omitted and/or differ across implementations, making comparisons between models biased. It is also common practice to compare setups by only training and evaluating on fixed intervals of the historical data whereas, in a real setting, models will likely be retrained continuously as new market data becomes available over time. Since the performance of RL agents fluctuates between different market periods and the effect that each of the different components of a setup has on the results can be difficult to ascertain, being able to easily implement different setups, train agents on them, and deploy monitoring protocols of agent behavior on novel data are necessary steps to successfully exploit RL-based Optimal Execution algorithms in real settings. However, the limited availability of documentation and code implementations of setups presents a barrier for entry.\\

This work aims to lower this barrier of entry by showcasing a modular RL framework that eases the implementation of different environments/simulation setups by separately implementing the specific simulation choices of each setup through different modules. Our work can be of interest to both practitioners and academics since it provides a basis that can vertebrate and serialize the implementation of all the aspects of RL Optimal Execution, allowing for a more uniform way to implement, benchmark and monitor different simulation setups.\\

In Section \ref{Section2 Modular Implementation}, we give a short introduction to RL and introduce the framework by describing each module and specifying the functionalities it should incorporate. We implement each module as a separate Python class and structure them as depicted in Figure \ref{Class Structure Graph}. We discuss the difficulties that the implementation of each individual module entails, the conflicts arising from their dependencies and their interactions with other modules.\\

In Section \ref{Section3 Setup Example} we showcase the usage of the modular implementation by implementing a setup intended to enhance a TWAP algorithm via RL, in the spirit of Bertsimas \& Lo \cite{Bertsimas1998OptimalCO} and Almgren \& Chriss \cite{almgren2000optimal}. The setup lets the RL agent decide the volume of limit orders submitted following the order submission schedule of a TWAP algorithm and imposes a bucket scheduling logic: At the beginning of execution, it divides the execution horizon in a number of periods/buckets and assigns volume to be executed proportionally to each bucket. At the end of each bucket, the remaining unexecuted volume is submitted as a market order. The reasoning behind following this kind of order scheduling logic, which is common practice in many algorithms, is that we minimize the market risk, and thus the probability of a very unfavorable execution. The execution of each limit order is simulated via iterating over snapshots of the limit order book (LOB) and rewards are calculated as the \$ improvement over the price achieved by the TWAP benchmark algorithm subject to the same bucket scheduling logic over the same time horizon. As already mentioned, these simulation choices make for a training environment that is different from a real market, so we discuss the explicit and implicit differences that our particular implementation induces in the simulation, and emphasize the importance of considering this phenomenon, the training-deployment distribution shift, as an essential aspect of Optimal Execution RL-based algorithm development.\\

Finally, in Section \ref{Section4 Experiments}, we showcase two different experiments. In the first experiment, we iteratively re-train and evaluate different models over a set of contiguous 5-day intervals of market data, mimicking how models may behave out of sample when being continuously retrained and then immediately deployed as new market data becomes available, thus providing a closer estimation of live model performance. Alternatively, in the second experiment, we investigate whether a model's out of sample performance decays over time, to estimate the consequences of not re-training a model.

\section{A Modular Implementation for RL Optimal Execution}\label{Section2 Modular Implementation}

\begin{figure}[h!]
    \centering
    \includegraphics[width=0.4\textwidth,center]{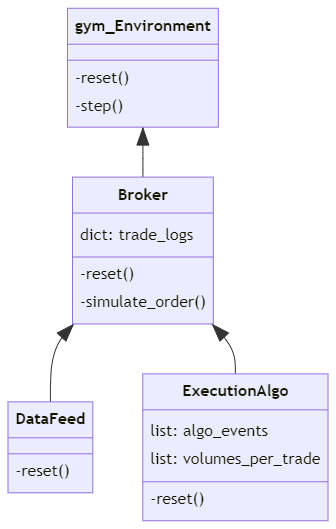}
    \caption{Modules Dependency Structure.}
    \label{Class Structure Graph}
\end{figure}

In this Section, after a brief introduction to RL, we introduce a modular framework that allows for the separate implementation of all the different functionalities and internal processes that an Optimal Execution RL environment needs to incorporate. The core modules we will introduce are the \code{Data Feed}, the \code{Broker} and the \code{Execution Algo}, see Table \ref{table:Modules Table} for a summarized description of their functionalities and Figure \ref{Class Structure Graph} for an illustration of their dependency structure.

\begin{table}[]
    \centering
    \begin{tabular}{c c c c}
    \toprule
    \textbf{Class Name} & \textbf{Functionalities} & \textbf{Required Methods}  & \textbf{Dependency} \\\toprule
    \code{Data Feed} & Data retrieval and pre-processing & \code{reset()} & \code{Broker} attribute \\\midrule
    \code{\texttt{Execution Algo}} & Execution Logic  & \code{reset()} & \code{Broker} attribute \\\midrule
    \code{Broker} & Order Execution Simulation & \code{reset()} \& \code{\texttt{simulate\_order()}} & environment attribute \\\midrule
    gym environment & RL logic & \code{reset()} \& \code{step()} & \textit{None} \\

    \bottomrule
    \end{tabular}
    \caption{Proposed modules of the Optimal Execution RL framework.}
    \label{table:Modules Table}
\end{table}

\subsection{Introduction}
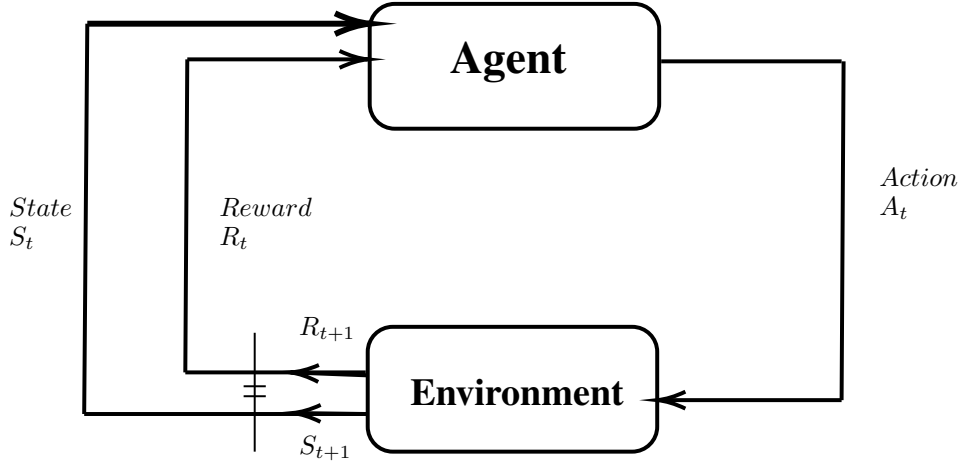
\begin{figure}
\centering

 
\tikzset{
pattern size/.store in=\mcSize, 
pattern size = 5pt,
pattern thickness/.store in=\mcThickness, 
pattern thickness = 0.3pt,
pattern radius/.store in=\mcRadius, 
pattern radius = 1pt}\makeatletter
\pgfutil@ifundefined{pgf@pattern@name@_bg8ezy9bl}{
\pgfdeclarepatternformonly[\mcThickness,\mcSize]{_bg8ezy9bl}
{\pgfqpoint{-\mcThickness}{-\mcThickness}}
{\pgfpoint{\mcSize}{\mcSize}}
{\pgfpoint{\mcSize}{\mcSize}}
{\pgfsetcolor{\tikz@pattern@color}
\pgfsetlinewidth{\mcThickness}
\pgfpathmoveto{\pgfpointorigin}
\pgfpathlineto{\pgfpoint{\mcSize}{0}}
\pgfpathmoveto{\pgfpointorigin}
\pgfpathlineto{\pgfpoint{0}{\mcSize}}
\pgfusepath{stroke}}}
\makeatother
\tikzset{every picture/.style={line width=0.75pt}} 

 
\tikzset{
pattern size/.store in=\mcSize, 
pattern size = 5pt,
pattern thickness/.store in=\mcThickness, 
pattern thickness = 0.3pt,
pattern radius/.store in=\mcRadius, 
pattern radius = 1pt}\makeatletter
\pgfutil@ifundefined{pgf@pattern@name@_drrntmntl}{
\pgfdeclarepatternformonly[\mcThickness,\mcSize]{_drrntmntl}
{\pgfqpoint{-\mcThickness}{-\mcThickness}}
{\pgfpoint{\mcSize}{\mcSize}}
{\pgfpoint{\mcSize}{\mcSize}}
{\pgfsetcolor{\tikz@pattern@color}
\pgfsetlinewidth{\mcThickness}
\pgfpathmoveto{\pgfpointorigin}
\pgfpathlineto{\pgfpoint{\mcSize}{0}}
\pgfpathmoveto{\pgfpointorigin}
\pgfpathlineto{\pgfpoint{0}{\mcSize}}
\pgfusepath{stroke}}}
\makeatother
\tikzset{every picture/.style={line width=0.75pt}} 

\begin{tikzpicture}[x=0.75pt,y=0.75pt,yscale=-1,xscale=1]

\draw  [line width=1.5]  (301,42.6) .. controls (301,35.64) and (306.64,30) .. (313.6,30) -- (434.4,30) .. controls (441.36,30) and (447,35.64) .. (447,42.6) -- (447,80.4) .. controls (447,87.36) and (441.36,93) .. (434.4,93) -- (313.6,93) .. controls (306.64,93) and (301,87.36) .. (301,80.4) -- cycle ;
\draw  [line width=1.5]  (300,205.6) .. controls (300,198.64) and (305.64,193) .. (312.6,193) -- (433.4,193) .. controls (440.36,193) and (446,198.64) .. (446,205.6) -- (446,243.4) .. controls (446,250.36) and (440.36,256) .. (433.4,256) -- (312.6,256) .. controls (305.64,256) and (300,250.36) .. (300,243.4) -- cycle ;
\draw [line width=1.5]    (448,59) -- (501,59) -- (539,59) ;
\draw [line width=1.5]    (538.99,230) -- (449,230) ;
\draw [shift={(446,230)}, rotate = 360] [color={rgb, 255:red, 0; green, 0; blue, 0 }  ][line width=1.5]    (14.21,-4.28) .. controls (9.04,-1.82) and (4.3,-0.39) .. (0,0) .. controls (4.3,0.39) and (9.04,1.82) .. (14.21,4.28)   ;
\draw [line width=1.5]    (540,59) -- (538.99,230) ;
\draw [line width=1.5]    (208,216) -- (261,216) -- (299,216) ;
\draw [line width=1.5]    (209,58) -- (297,58) ;
\draw [shift={(300,58)}, rotate = 180] [color={rgb, 255:red, 0; green, 0; blue, 0 }  ][line width=1.5]    (14.21,-4.28) .. controls (9.04,-1.82) and (4.3,-0.39) .. (0,0) .. controls (4.3,0.39) and (9.04,1.82) .. (14.21,4.28)   ;
\draw [line width=1.5]    (209,58) -- (208,216) ;
\draw [line width=1.5]    (157,237) -- (239.96,237) -- (299.43,237) ;
\draw [line width=2.25]    (158.57,40) -- (297,40) ;
\draw [shift={(301,40)}, rotate = 180] [color={rgb, 255:red, 0; green, 0; blue, 0 }  ][line width=2.25]    (17.49,-5.26) .. controls (11.12,-2.23) and (5.29,-0.48) .. (0,0) .. controls (5.29,0.48) and (11.12,2.23) .. (17.49,5.26)   ;
\draw [line width=1.5]    (158.57,40) -- (157,237) ;
\draw [line width=1.5]    (300,217.5) -- (264,216.12) ;
\draw [shift={(261,216)}, rotate = 2.2] [color={rgb, 255:red, 0; green, 0; blue, 0 }  ][line width=1.5]    (14.21,-4.28) .. controls (9.04,-1.82) and (4.3,-0.39) .. (0,0) .. controls (4.3,0.39) and (9.04,1.82) .. (14.21,4.28)   ;
\draw [line width=1.5]    (299.43,237) -- (265,236.54) ;
\draw [shift={(262,236.5)}, rotate = 0.77] [color={rgb, 255:red, 0; green, 0; blue, 0 }  ][line width=1.5]    (14.21,-4.28) .. controls (9.04,-1.82) and (4.3,-0.39) .. (0,0) .. controls (4.3,0.39) and (9.04,1.82) .. (14.21,4.28)   ;
\draw [pattern=_drrntmntl,pattern size=6pt,pattern thickness=0.75pt,pattern radius=0pt, pattern color={rgb, 255:red, 0; green, 0; blue, 0}]   (243,196) -- (243,256) ;
\draw [shift={(243,223)}, rotate = 270] [color={rgb, 255:red, 0; green, 0; blue, 0 }  ][line width=0.75]    (0,5.59) -- (0,-5.59)(-5.03,5.59) -- (-5.03,-5.59)   ;

\draw (340,47.5) node [anchor=north west][inner sep=0.75pt]  [font=\LARGE] [align=left] {\textbf{{\LARGE Agent}}};
\draw (320,217.5) node [anchor=north west][inner sep=0.75pt]  [font=\Large] [align=left] {\textbf{Environment}};
\draw (550,110.4) node [anchor=north west][inner sep=0.75pt]    {$ \begin{array}{l}
Action\\
A_{t}
\end{array}$};
\draw (217,125.4) node [anchor=north west][inner sep=0.75pt]    {$ \begin{array}{l}
Reward\\
R_{t}
\end{array}$};
\draw (111,125.4) node [anchor=north west][inner sep=0.75pt]    {$ \begin{array}{l}
State\ \\
S_{t}
\end{array}$};
\draw (264,186.4) node [anchor=north west][inner sep=0.75pt]    {$R_{t+1}$};
\draw (264,246.4) node [anchor=north west][inner sep=0.75pt]    {$S_{t+1}$};

\end{tikzpicture}
\caption{Components and flows of the RL framework.} \label{fig:RLDiagram}
\end{figure}

Reinforcement Learning is a Machine Learning framework in which the objective is to learn rules for iterative decision making so that a specific numerical quantity, called a reward, is optimized. The algorithm tasked with obtaining rules is called the \textit{agent}; it receives as input the state of the world $S_{t}$ via another algorithm, called the \textit{environment}. The agent maps states to actions via a special function called the \textit{policy}. Given an action $A_t$ taken by the agent, the environment produces an updated value of the reward $R_{t+1}$ as well as a new state of the world $S_{t+1}$, see Figure \ref{fig:RLDiagram} for an illustration. The main difference between RL and supervised learning problems is that the new states of the world may be affected by the agent's actions (i.e. there is interaction between the components of the problem).\\

More formally, the RL framework is characterized as a \textit{Markov Decision Process} (MDP). The MDP framework describes the agent–environment interaction through the three signals mentioned above, the actions of the agent $\{A_t\}_t$, the states of the environment $\{S_t\}_t$ and the  rewards $\{R_t\}_t$. An MDP is described by a sequence of discrete time steps $t_0, ..., T$ and a tuple $\left\{\mathcal{S}, \mathcal{A}(S), \mathbb{P}\left(S^{\prime} \mid S, A\right), R, \gamma\right\}$ in which:

\begin{enumerate}
    \item $\mathcal{S}$ is the set of states so that each $S_t \in \mathcal{S}$.
    \item $\mathcal{A}(S)$ defines possible actions $A_t \in \mathcal{A}(S)$ that can be taken in a state $S_t = S$.
    \item $\mathbb{P}\left(S^{\prime} \mid S_{t-1}, A_{t-1}\right)$ represent the transition probabilities of states $S^{\prime} \in \mathcal{S}$ given action $A_{t-1}$ is chosen at a previous state $S_{t-1}$.
    \item $R$ and $\gamma$ represent the set of all possible rewards and the discount factor, which determines how much more desirable are immediate rather than latter rewards.
\end{enumerate}

The transition probabilities have to fulfill the \textit{Markov Property}: Given $\mathbb{P}\left(S_{t} \mid S_{0: t-1}\right) = \sum_{A_t} \mathbb{P}(A_t) \mathbb{P}\left(S_{t} \mid S_{0: t-1}, A_t\right)$ (in integral form if the action space is continuous), we have that: 

\begin{equation}
\mathbb{P}\left(S_{t} \mid S_{0: t-1}\right)=\mathbb{P}\left(S_{t} \mid S_{t-1}\right)
\end{equation}

That is, the probabilities of arriving at any state $S_t$ depend exclusively on the previous state $S_{t-1}$. The goal of an agent is to find a policy $\pi$ that maximizes the expected discounted cumulative reward:

\begin{equation}
\argmax_{\pi}\mathbb{E}^{\pi}\left[\sum_{t=0}^{T-1} \gamma^{t} R_{t}\left(S_{t}, A_{t}\right)\right]
\end{equation}
for which there exist multiple optimization methods \cite{SuttonBarto2018, dixon2020machine}.

\subsection{The \code{Data Feed} Class}\label{DataFeedClassModule}

Datasets of High-Frequency (HF) market data can come in many different formats, with different levels of granularity, that allow for different simulation setups. They can be broadly categorized (although there are exceptions in specific markets) into what is commonly referred to as "Levels":
\begin{enumerate}
    \item \textbf{Level 1 data:} Consists of records of traded prices at different timestamps. It may include the very first levels of the LOB, the best bid/ask and their respective volumes as well.
    \item \textbf{Level 2 data:} Extends Level 1 data by including deeper levels of the LOB, that is, the price levels greater/lower than the lowest ask/highest bid respectively and their volumes. This is usually referred to as "snapshots" of the LOB. They may have a fixed size or change at each timestamp. Often price levels very far away from the best bid/ask are not recorded whereas in a real market this can be a possibility.
    
    \item \textbf{Level 3 data:} Consists of the sequence of limit order placements, deletions and executions. It is the most granular out of the three types, since the other two can be reconstructed from it, and can lead to large datasets because of this.
\end{enumerate}

Besides the data format, the regularity at which each datapoint is recorded also plays a very big role later on in the simulation. It can range from having some form of HF data sampled at a constant rate to having a datapoint for each order placement, deletion, execution, and other market events (volatility halts, circuit breakers etc.). Data accessibility can vary between different markets, and more granular data is often either pay-walled or unavailable. Thus, being able to process different data formats is a necessity. Because of this, we code \code{Data Feeds} as Python Subclasses, each with different methods adapted to their specific data format.\\

Each \code{Data Feed} Subclass needs to be able to retrieve data iteratively in order to allow the \code{Broker} to produce observations for the agent in the environment's \code{step} function. If we are dealing with a dataset with entries ordered by timestamps, one of the easiest ways to do this is to have an internal state that represents the current time of the simulation and signals to the next datapoint to be retrieved from the data. One thing to keep in mind is that the timestamps of order placements may not necessarily align with the timestamps of the data entries, and thus it becomes necessary to incorporate logic in the \code{Data Feed} that finds the closest datapoint after a given timestamp. In the same way the environment incorporates a \code{reset} function called at the beginning of each Episode, each \code{Data Feed} Subclass needs a \code{reset} function that will be called simultaneously and reconfigures its internal state for the next execution.\\

In addition, some setups simulate synthetic markets through artificial data \cite{MarketSimGAN,MultiAgentDeepRLLiqStrats, DeepExeRLTradingBeating}. All of this logic can be accommodated as methods of a \code{Data Feed} subclass in the same fashion as we do with historical data and the internal state modified to suit the needs of the specific algorithm.
\vspace{.25cm}

\subsection{The \code{Broker} Class}

An RL agent needs to be able to interact with its environment, the market under an Optimal Execution setting, during the training process. In order to implement this feedback mechanism between the market and the agent's actions, translated as order submissions, we need to be able to simulate order executions. We call this module the \code{Broker} Class since it is responsible of placing orders into the market, simulating their execution and keeping logs of the trades carried out over the entire trading horizon.\\

When presented with an order, the \code{Broker} has to be able to simulate its execution through an algorithm of choice, that may be different for each \code{Execution Algo}, by interacting with its \code{Data Feed}. Each algorithm may react differently to partial executions, thus the Broker needs to incorporate as methods each of the different ways \code{Execution Algos} may act mid-execution. Examples of these choices are submitting all orders as market orders, placing limit orders but deleting them and replacing them with a price closer to the best bid/ask after they have remained in the LOB for a given time, combining market and limit orders simultaneously at each order submission etc.\\

\begin{figure}[h!]
    \includegraphics[width=1.075\textwidth,center]{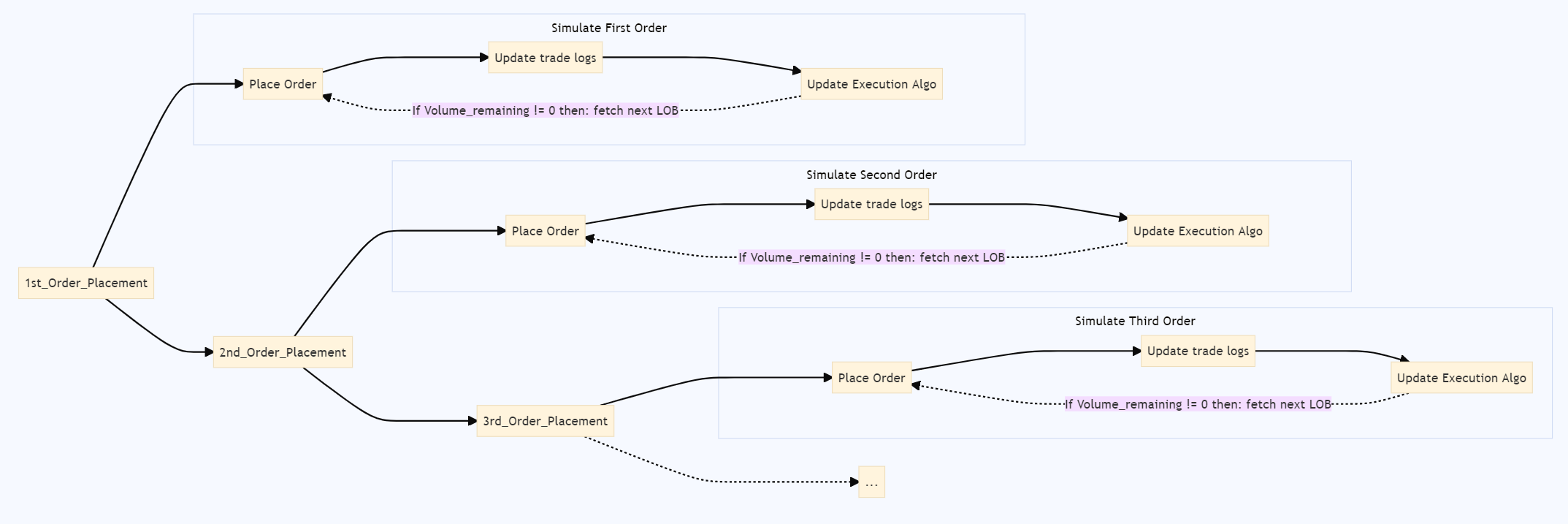}
    \caption{Typical \code{Broker} simulation workflow. Each order placement and its corresponding simulation represent a step of the RL agent.}
    \label{Broker Execution Diagram}
\end{figure}

The \code{Broker} needs to be able to process orders originating from different \code{Execution Algos}, records like the trade logs, the LOB histories and others need to be implemented as dictionaries whose keys represent the \code{Execution Algos} being simulated. Since the \code{Broker} is responsible of keeping track of all the execution related variables at any point of the simulation, it needs to be able to modify the internal state of the corresponding \code{Execution Algos}, so setting them as \code{Broker} attributes is the most straightforward way of implementing this dependency. See Figure \ref{Broker Execution Diagram} for an illustration of a typical \code{Broker} simulation workflow.\\

Just as the environment and the \code{Data Feed}, the \code{Broker} needs to include a \code{reset} function that, when called at the beginning of each episode, will clear all its internal logs, reset its \code{Data Feed} according to the new execution starting time and reset its \code{Execution Algos} according to the new execution parameters.

\vspace{.25cm}

\subsection{The \code{Execution Algo} Class}

Each Execution Algorithm divides a transaction into child orders across the execution interval according to some internal logic. The order submission schedule and each order's volume may be decided a priori (like a TWAP algorithm), may be determined via the algorithm's internal state and/or market conditions during the execution (like a RL algorithm) or may incorporate a combination of both (like an algorithm that incorporates a buckets schedule).\\

The \code{Execution Algo} class encodes a representation of each execution algorithm consisting of its actions indexed by their timestamps (\code{\texttt{algo\_events}}), the variables involved in each action (volume, type of order etc.), as well as variables involved in the particular logic of the algorithm and an internal state containing the execution parameters such as the trading direction, the total volume to execute or the volume left to execute.

Just as the two classes before, each \code{Execution Algo} needs to also include a \code{reset} function that, when called at the beginning of each episode, will reset all the internal variables involved in the algorithm's logic and reset its execution parameters.

\vspace{.25cm}

\subsection{The \code{gym} environment Class}

The environment is the core of the RL methodology, it is tasked with processing the agent's actions, calculating the corresponding rewards and producing subsequent states.  An environment can be implemented through the specification of two required methods, the \code{reset} \& \code{step} functions that carry out the tasks previously described and specify the dimensions \& types (continuous or discrete) of the Observation and Action Spaces that the agent will receive Observations and be allowed to draw actions from respectively.

\subsubsection{Executions as Episodes}\label{Execution as Episodes Section}

In a RL Optimal Execution setting we consider each Execution to be a different training Episode. At the start of each Episode, the \code{reset} function is called and the execution parameters are randomized. All setups incorporate the following execution parameters:

\begin{enumerate}
    \item \code{\texttt{start\_time}}: Timestamp of the beginning of the execution.
    
    \item \code{\texttt{exec\_time}}: Duration of the current execution.

    \item \code{\texttt{trade\_direction}}: Whether the transaction is a Buy/Sell.
    
    \item \code{volume}: The number of units of the asset to Buy/Sell in the Execution.
\end{enumerate}

In addition to setup-specific parameters. Additionally, the \code{reset} function has to reset the \code{reward} and \code{info} attributes, as well as the execution-specific attributes of the environment, that is, the \code{Broker} and through its reset function its \code{Execution Algos} and its \code{Data Feed}. It also needs to produce the first state for the agent to process at the beginning of each Episode.

\subsubsection{Order placements as steps}

The \code{step} function is called when the agent produces an action after being presented with a state of the environment. In a RL Optimal Execution setting, it needs to incorporate the necessary logic to accomplish the following: 

\begin{enumerate}
    \item Translate the agent's actions into order submissions, determining its submission type, order type and volume through interacting with the RL \code{Execution Algo}.
    
    \item Simulate via the \code{Broker} the Execution of the orders submitted by all the \code{Execution Algos} involved in the reward calculation up until the timestamp of the next agent's action.
    
    \item Calculate rewards at specific timestamps of the execution that depend on the setup. Some setups reward the agent at every order submission, some at specific timestamps, some only at the end of the execution etc.
\end{enumerate}

\begin{figure}[h!]
    \includegraphics[width=0.75\textwidth,center]{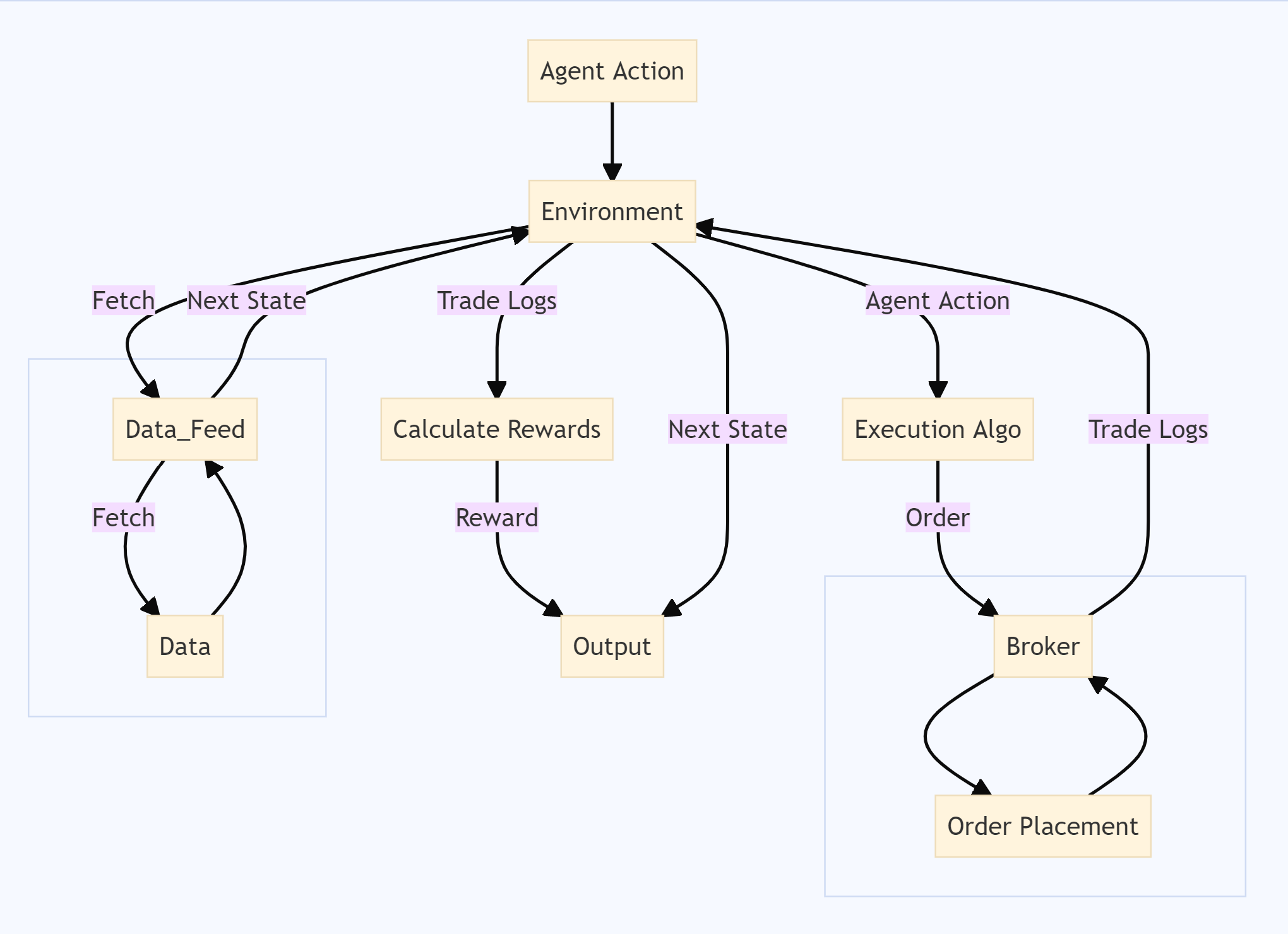}
    \caption{Typical workflow for the environment \code{step} function.}
    \label{Environment Step Diagram}
\end{figure}

Each environment needs to calculate rewards via interacting with the \code{\texttt{trade\_logs}} of the \code{Broker}. Different reward calculations can incentivize different aspects of the RL Execution, not only the average price achieved, but also the speed, the slippage incurred, the market vs limit order volume ratio etc. As in all RL setups, the \code{step} function needs to also produce the subsequent \code{state} the agent will be presented with via the \code{Data Feed} and ascertain whether the current episode is over. See Figure \ref{Environment Step Diagram} for an illustration of the workflow of the \code{step} function.

\subsection{UnitTests}

For the most part the implementation of each of the modules described above can be rather intuitive, but given that our aim is to implement numerous different setups dealing with different data sources and performing different calculations, it is advisable to implement a series of Unit Tests on each implementation. Common aspects to check for include:

\begin{enumerate}
    \item \textbf{Reproducibility:} Given two \code{Execution Algos} with equal representations (meaning they have equal \code{\texttt{algo\_events}} and \code{\texttt{volumes\_per\_trade}}), they should produce \textbf{identical} \code{\texttt{trade\_logs}} when simulated by identical \code{Brokers} (meaning they share all the attributes and thier \code{Data Feeds} have been reset to the same timestamp). Failing to pass this test can mean that the \code{Data Feed} is permanently modifying historical data and/or that the \code{reset} functions of some modules are not properly implemented. 
    \item \textbf{Duplicity:} Each datapoint from the historical data should be interacted with strictly \textbf{once}. Failing to pass this test can mean that the \code{Data Feed}, or modules that make use of it, are incorrectly iterating through the historical data.
    \item  \textbf{Rounding Issues:} At the end of each Execution, it is important to verify that the executed volume accounted for in the \code{\texttt{trade\_logs}} is equal to the initial \code{volume} of the Execution. All calculations that involve market variables should be done with Python \code{Decimals} rather than floats, in order to avoid rounding errors and keep the rounding to the minimum tick size of the given market consistent across the whole simulation. Failing to pass this test can mean that the execution of some orders is not correctly implemented (for example, unlike in a real market, submitting a market order vs a LOB snapshot does not guarantee a full execution, since the LOB may not have enough volume to fully consume the order). 

\end{enumerate}

For efficiency reasons, all the UnitTests should be implemented outside the implementation of the environment, since they will introduce overhead to the RL training. As a final note, it is important to implement each of the modules described in this section with efficiency in mind, trying to avoid unnecessary computations as much as possible and taking advantage of any achievable efficiency improvements. An inefficient implementation of any of the components can effectively make the training of a setup unmanageable in practice. 

\section{Setup Example}\label{Section3 Setup Example}

In this section, we present an example implentation of an Optimal Execution setup through our modular framework. We share the Python code implementation in \footnote{\url{https://github.com/FernandoDeMeer/RL_Optimal_Execution}}. The setup we choose to showcase aims to improve a TWAP benchmark algorithm via RL: we only allow the agent to submit orders at the same timestamps as the TWAP benchmark and to choose only the order volumes as 80\%,100\% or 120\% of the TWAP volumes through its actions. The prices of limit orders are set to be 1 tick away from the best bid/ask and simulated according to a procedure we will shortly present. Because of this, the RL agent is subject to a series of restrictions that we implement via specific subclasses of the modules described in Section \ref{Section2 Modular Implementation}.

\subsection{\code{Data Feed} Example}

We use a dataset of Level 2 HF data consisting of snapshots of the LOB of BTC/USDT futures of Binance, a crypto-currency exchange, up to 10 price levels away from the best bid/ask for each side (adding up to 20 levels every snapshot)\footnote[1]{See: \url{https://github.com/binance/binance-public-data/tree/master/Futures_Order_Book_Download}} and recorded every 100 ms from 01/06/2021 to 20/06/2021. We have different files for each day, therefore we partition the training and evaluation periods by days.

The snapshots are indexed by timestamps and thus make for a straightforward implementation of the \code{Historical Data Feed} SubClass. The SubClass only needs to:

\begin{enumerate}
    \item \textbf{Load} the dataset from source files and store it as an attribute.
    \item \textbf{Implement}  a \code{reset} method that locates the closest LOB to a given timestamp and records its position in the dataset.
    \item \textbf{Fetch past LOBs} at a given timestamp and iterate through the data in order to allow the \code{Broker} to simulate order executions and build states for environment.
\end{enumerate}

\begin{algorithm}
\caption{Simulation of a Limit Buy Order Execution}
\label{LimitOrderExeAlgo}
\hspace*{\algorithmicindent} \textbf{Input:} Limit Order $\mathcal{O}_l$, \code{Data Feed} $\mathbb{D}$, \code{Execution Algo} $\mathbb{A}$, current timestamp $t$, market \code{\texttt{tick\_size}}\\
\hspace*{\algorithmicindent} \textbf{Output:} Execution of order $\mathcal{O}_l$ recorded in  \code{Broker.\texttt{trade\_logs}}
\begin{algorithmic}[1]
\While{$\mathcal{O}_l\text{\code{.volume}} > 0 \And t < \mathbb{A}.\code{\texttt{next\_event.time}}$ }:
\State $\mathcal{O}_l$\code{.price} = $\mathbb{D}.\code{\texttt{current\_LOB.get\_best\_bid()}} - \code{\texttt{tick\_size}}$
\State Update $\mathbb{D}$ $\Rightarrow$ $\mathbb{D}.\code{\texttt{next\_LOB()}}$
\State Update $t$ $\Rightarrow$ $ t =\mathbb{D}.\code{\texttt{current\_LOB.time}} $
\State $ \code{\texttt{trade\_log}} = \code{\texttt{place\_order}}(\mathcal{O}_l,  \mathbb{D}.\code{\texttt{current\_LOB}})$,  append to  \code{Broker.\texttt{trade\_logs}}
\If{$ \code{\texttt{trade\_log.message}} == trade$}
\State Update $\mathcal{O}_l\text{\code{.volume}}$ $\Rightarrow$ $\mathcal{O}_l\text{\code{.volume}} \minuseq \code{\texttt{trade\_log.volume}}$
\State Update $\mathbb{A}\text{\code{.\texttt{remaining\_volume}}}$ $\Rightarrow$ $\mathbb{A}\text{\code{.\texttt{remaining\_volume}}} \minuseq \code{\texttt{trade\_log.volume}}$
\If{$\mathcal{O}_l\text{\code{.volume}} = 0$}
\State \textbf{break}
\EndIf
\EndIf
\If{$t \geq \mathbb{A}.\code{\texttt{next\_event.time}} \And \mathbb{A}.\code{\texttt{next\_event}} == \code{\texttt{limit\_order}}$}
\State $\mathbb{A}.\code{\texttt{next\_event.volume}}\pluseq \mathcal{O}_l\text{\code{.volume}}$
\Else
\If{ $t \geq \mathbb{A}.\code{\texttt{next\_event.time}}$}
\State Submit a market order of $\mathbb{A}\text{\code{.\texttt{remaining\_volume}}}$
\EndIf
\EndIf
\EndWhile
\end{algorithmic}
\end{algorithm}

\begin{algorithm}
\caption{Simulation of a Market Buy Order Execution}
\label{MarketOrderExeAlgo}
\hspace*{\algorithmicindent} \textbf{Input:} Market Order $\mathcal{O}_m$, \code{Data Feed} $\mathbb{D}$, \code{Execution Algo} $\mathbb{A}$, current timestamp $t$\\
\hspace*{\algorithmicindent} \textbf{Output:} Execution of order $\mathcal{O}_m$ recorded in  \code{Broker.\texttt{trade\_logs}}
\begin{algorithmic}[1]
\While{$\mathcal{O}_m\text{\code{.volume}} > 0$}
\State Update $\mathbb{D}$ $\Rightarrow$ $\mathbb{D}.\code{\texttt{next\_LOB()}}$
\State Update $t$ $\Rightarrow$ $ t =\mathbb{D}.\code{\texttt{current\_LOB.time}} $
\State $ \code{\texttt{trade\_log}} = \code{\texttt{place\_order}}(\mathcal{O}_m,  \mathbb{D}.\code{\texttt{current\_LOB}})$,  append to  \code{Broker.\texttt{trade\_logs}}
\State Update $\mathcal{O}_m\text{\code{.volume}}$ $\Rightarrow$ $\mathcal{O}_m\text{\code{.volume}} \minuseq \code{\texttt{trade\_log.volume}}$
\State Update $\mathbb{A}\text{\code{.\texttt{remaining\_volume}}}$ $\Rightarrow$ $\mathbb{A}\text{\code{.\texttt{remaining\_volume}}} \minuseq \code{\texttt{trade\_log.volume}}$
\EndWhile
\end{algorithmic}
\end{algorithm}

\subsection{\code{Broker} Example}

As discussed in Section \ref{DataFeedClassModule}, there exists a spectrum of HF datasets with different granularities. More granular datasets allow for simulations of order executions that incorporate more market information. In our setup, we have Level 2 Data in the form of LOB snapshots and thus can only use this information to simulate the market response to order placements. When using order book level data to simulate executions, it is advisable to employ a LOB Class implementation that efficiently carries out the basic interactions between order books and orders (placements, deletions and executions). In our case, we use \cite{OrderBookGithubRepo} which implements price levels as linked lists (by time of arrival) and the Bids and Asks of each LOB as 2 red-black trees.\\

Algorithm \ref{LimitOrderExeAlgo} contains the pseudo-code of our chosen simulation of the execution of limit buy orders. The volume of the limit order is decided by the RL agent and the price is chosen as the best bid of the current LOB -1 tick (+1 in case of a sell order). Then, the order is matched against the next LOB snapshot. We record the placement and, if it leads to a trade (when the market moves downwards w.r.t to the last LOB), we update the remaining volume of the \code{Execution Algo} submitting the order accordingly and repeat the process with the subsequent LOBs until either the volume of the order is fully consumed or we reach the next order placement, in which case we add the remaining volume to that of the new limit order, or reach the end of the \code{Bucket}, which submits the remaining volume as a market order. In a real setting, this update of the price of the standing order every snapshot would require cancelling it and resubmitting a new order, but unlike in our simulation, the order could get (partially) consumed between snapshots. This simulation approach executes only after favorable market movements and thus incorporates the delay in execution that limit orders entail.\\

Algorithm \ref{MarketOrderExeAlgo} contains the pseudo-code of our chosen simulation of the execution of market buy orders. Unlike in a real market, when placing a market order in a LOB snapshot, the market order might not be fully executed because the LOB might lack enough volume on the corresponding side to do so, since the snapshots carry only the 10 closest price levels to the best bid/ask. Because of this, it also becomes necessary to iterate through LOBs in the execution of market buy orders.\\

The environment of our setup calculates rewards based on the outperformance of the RL agent w.r.t a TWAP algorithm following the same order submission schedule. Because of this, we carry out the two previous Algorithms in parallel in our simulation, that is, both the RL and the benchmark TWAP algorithm have access to the same LOBs when placing orders; their actions do not affect the other in any way. This is justified because we are not interested in having both Algorithms compete, but rather just obtaining a meaningful signal to train the RL Algo with.   \\

By simulating the execution of limit and market orders via Algorithms \ref{LimitOrderExeAlgo} and \ref{MarketOrderExeAlgo} we are inducing the following phenomena, which distance the behaviour of our simulation from that of a real market:

\begin{enumerate}
    \item \textbf{Underestimating volume}: Since we have only LOB snapshots, we are missing all the order placements and executions that took place between snapshots and thus may be underestimating the volume available in the market.
    \item \textbf{Overestimating volume}: We have no way of knowing whether the volumes appearing at two LOBs with different timestamps originate from the same order placements or not, thus we might be executing the same volume more than once and thus may be overestimating the volume available in the market.
    \item  \textbf{Lack of LOB Depth}: since the LOB snapshots we have available have always the same size, we have no knowledge of the orders placed 10 levels away from the best bid/ask. This makes it so that we ignore volumes placed at very unfavourable price levels for our agent, specially for the execution of market orders.
    \item \textbf{No market impact}: Since we are simulating via a static historical dataset, we are not currently incorporating into the simulation the reaction that other market participants would have to the actions of our agent, which would shift the behaviour of a real market.
    \item \textbf{Transaction fees}: We are not taking into account the exchange fees incurred in by our transactions. In Optimal Execution, fees may have a big impact on the performance of an algorithm, specially if they're charged on a per-order basis such as market access fees.
    \item \textbf{Latency}: Contrary to most setups, we do introduce a delay for our agent's actions, since order submissions are executed against future LOBs rather than the one corresponding to the observation that prompted the order submission. However, we still assume that our submission will take less than 100ms to reach the market, which may or mat not be realistic depending on the market at hand. 
\end{enumerate}

Issues 1-3 can be improved via having access to more granular datasets (the sequence of order placements, deletions and executions specifically) in settings where they are available. Issues 5 \& 6 require setup and market-specific knowledge to improve but they are relatively easy to work around. Incorporating Issue 4 requires a model-based approach and is the focus of an active branch of the financial simulations literature, see \cite{Naritomi2020DataAO, GeneratingRealistOrdStreams2020,MarketSimGAN, Hirano2022PolicyGS}.

\subsection{\code{Execution Algo} Example}

\begin{figure}[!htb]
    \includegraphics[width=0.7\textwidth,center]{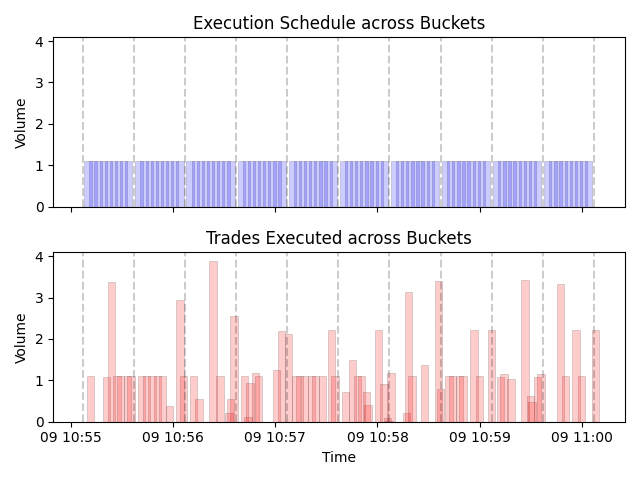}
    \caption{TWAP order submission schedule vs  executed trades following Algorithms \ref{LimitOrderExeAlgo} and \ref{MarketOrderExeAlgo}.}
    \label{TWAP_Execution_Diagram}
\end{figure}

We implement both the RL and the benchmark TWAP Algorithm as \code{Execution Algos}. The TWAP is implemented via an order submission schedule that divides the execution interval into segments referred to as \code{Buckets}, which are assigned volume proportionally according to their duration. Each bucket populates the \code{\texttt{algo\_events}} of the \code{Execution Algo} with the timestamps of a given number of equally distributed \code{\texttt{limit\_orders}} along its duration and adds a last timestamp to represent the \code{\texttt{bucket\_bound}} (which signals the submission of all the remaining bucket volume as a market order). See Figure \ref{TWAP_Execution_Diagram} for an illustration of this order submission schedule.\\

Since both types of events represent order submissions and the volumes of the market orders depend on the remaining volume (which depends on market movements), we only need to keep track of the limit order volumes through a list we call \code{\texttt{volumes\_per\_trade}} and the list of either the \code{\texttt{limit\_orders}}' or the \code{\texttt{bucket\_bounds}}' timestamps which allows us to check whether each event leads to a limit order or a market order submission. The RL Algorithm is implemented through inheriting all the attributes of the TWAP (since they share the order submission schedule) and modifies during training its \code{\texttt{volumes\_per\_trade}} list, depending on the agents actions. 

\begin{figure}[!htb]
    \includegraphics[width=0.7\textwidth,center]{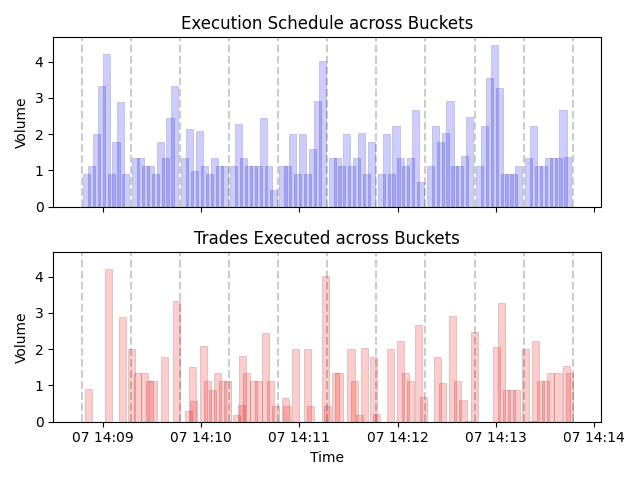}
    \caption{RL order submission schedule vs executed trades  following Algorithms \ref{LimitOrderExeAlgo} and \ref{MarketOrderExeAlgo}. Keep in mind that the RL agent decides its schedule over the course of training by adding volume to running orders, unlike the TWAP schedule which is decided a-priori without taking market conditions into account.}
    \label{RL_Execution_Diagram}
\end{figure}

\subsection{ \code{gym} environment Example}

We implement the environment of our setup via OpenAI's \code{gym} library \cite{OpenAIGymPaper2016}, by specifying its \code{reset} and \code{step} functions and a number of other submethods that implement specific functionalities. We choose a Discrete action space of size $n=3$ for the agent, representing the choice of submitting an additional volume to the limit order being currently executed (see Algorithm \ref{LimitOrderExeAlgo}) $\{0.8, 1, 1.2\} * \text{TWAP}_{Volume}$ for actions $\{0,1,2\}$ respectively. We also choose an Observation space of size $n=102$ consisting of a concatenation of the volumes and normalized prices of the 10 closest price levels in both the bids \& asks belonging to the past 5 LOB snapshots at the current timestamp (prices are normalized w.r.t to the mid-price of the latest observation) as well as two internal variables, the $\%$ of the bucket volume left to execute and the number of orders remaining inside the current bucket. In addition, the environment is initiated with a \code{Broker} as an attribute which in turn incorporates a \code{Data Feed} as described in previous sections.

\subsubsection{\code{reset} function Example}

In addition to resetting the attributes already mentioned in Section \ref{Execution as Episodes Section}, our setup \code{reset} function has to sample TWAP-specific attributes such as the \code{\texttt{no\_of\_slices}}, \code{\texttt{rand\_bucket\_bounds\_width}}, \code{\texttt{bucket\_func}} and \code{\texttt{delete\_vol}}, which specify different aspects of the order submission schedule, reset the, \code{\texttt{state\_idx}}, \code{\texttt{order\_idx}} and \code{\texttt{bucket\_idx}} attributes that represent internal indexes of the environment used to contextualize specific order submissions (such as market orders at bucket bounds), declare the new RL and TWAP benchmark \code{Execution Algos} with the attributes resampled above and build the first observation of the Episode for the agent.

\subsubsection{\code{step} function Example}

The \code{step} function of our setup implements the logic that translates the agent actions into order submissions via interacting with the \code{Execution Algo}, simulates order submissions until the next \code{\texttt{algo\_event}} via the \code{Broker} and Algorithms \ref{LimitOrderExeAlgo} \& \ref{MarketOrderExeAlgo}, creates new states by interacting with the \code{Data Feed} and calculates rewards at the end of each bucket as the total \$ difference, weighted by the volume executed, between the price achieved by the RL Algorithm vs that of the TWAP Benchmark:

\begin{equation}
    R_t(S_t,A_t) = \mathbbm{1}_{t\in T^*}(\text{RL}_{VWAP}-\text{TWAP}_{VWAP})
\end{equation}
In the case of selling and with opposite sign in the case of buying, where $T^* = \{t_1,t_2,...,t_N | t_i \text{ is a bucket bound}\}$ is the set of all the bucket bounds timestamps. The function follows the same workflow as Figure \ref{Environment Step Diagram}. See Figure \ref{RL_Execution_Diagram} for an illustration of an order submission schedule decided by the RL agent.

\section{Experiments}\label{Section4 Experiments}

In this section, we showcase two different experiments aimed at training agents and estimating their out of sample performance, carried out under the setup described in Section \ref{Section3 Setup Example}. In the first, we compare the out of sample performance of two agents continuously retrained over a set of 5-day periods in order to emulate an iterative training procedure that incorporates new market data as it becomes available. In the second, we investigate whether a model's out of sample performance decays over time if it is not continuously retrained. We use Ray's RLlib \cite{RLLibPaper2017,RayPaper2017}, an open-source Python library, to implement our experiments.

\begin{figure}[!tbp]
  \centering
  \begin{minipage}[b]{0.4\textwidth}
    \includegraphics[width=0.7\textwidth,center]{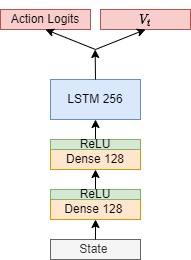}
    \caption{LSTM agent Architecture.}
    \label{LSTM Agent}
  \end{minipage}
  \hfill
  \begin{minipage}[b]{0.4\textwidth}
    \includegraphics[width=0.7\textwidth,center]{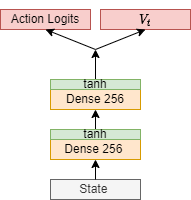}
    \caption{FCN agent Architecture.}
    \label{FCN Agent}
  \end{minipage}
\end{figure}

\subsection{Iterative Training Experiment}

We compare two different agents: the first architecture incorporates two Dense layers of 128 units each with ReLU activations followed by a LSTM layer with 256 units and two output dense layers of dimensions 3 and 1 for the action logits $a_t$ and the value function $V_t$ respectively. We refer to this architecture as the LSTM agent (see Figure \ref{LSTM Agent} for reference). The second architecture incorporates two Dense layers of 256 units each with tanh activations and the same output layers as before. We refer to this architecture as the FCN agent (see Figure \ref{FCN Agent} for reference). We train both agents via the PPO Algorithm \cite{2017arXiv170706347SPPOpaper}, one of the most popular policy-gradient algorithms, due to its state-of-the-art performance in numerous benchmarks and sample efficiency. We task both agents with buying 100 BTC over a 5-minute window and benchmark them against a TWAP that divides the execution interval into 10 buckets of 30s each and submits 9 limit orders in each bucket. Each agent is able to choose the volumes of its orders (but not the prices and type) as described in Section \ref{Section3 Setup Example}.\\

\begin{figure}[!tbp]
    \includegraphics[width=0.7\textwidth,center]{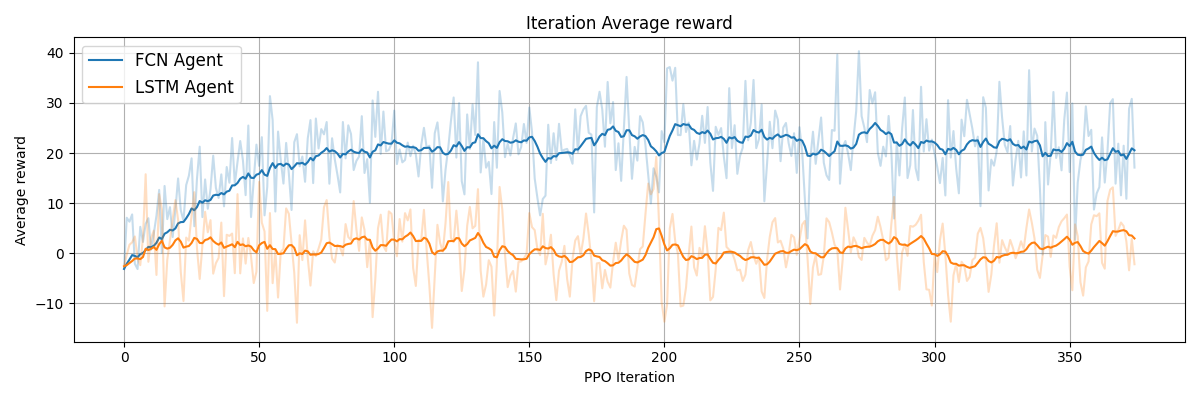}
    \caption{Smoothed Average Reward per PPO Iteration of both agents over the course of training, the FCN agent is plotted in blue and the LSTM agent in orange.}
    \label{Average reward plot}
\end{figure}

We train both agents over 3 different training periods, from 01/06/2021 to 05/06/2021, from 06/06/2021 to 10/06/2021 and from 11/06/2021 to 16/06/2021 and evaluate them over the 5 days following each period. For the second and third training periods, we start training with the best agent from the previous period. We train each agent over each training period for 125 PPO iterations, which account to approx. 1M steps of the environment, see Figure \ref{Average reward plot} for a plot of the Average Reward per Iteration over the course of training. The FCN agent manages to gradually improve its Average Reward over the first training instance (first 125 Iterations) and then plateaus from then onward, while the LSTM does not achieve the same improvement and stays with an average reward centered around 0 during the whole training process.\\

After training, we evaluate each model over its corresponding evaluation period for $10^4$ Episodes with the same execution parameters. The results are summarized in Table \ref{table:Iterative Performance Table}. We can observe that the LSTM agent barely outperforms the FCN over the first training period, but underperforms it more significantly over the other two. In addition, the FCN agent progressively improves its out of sample performance over the three periods after being trained with more data, while the LSTM agent does not. Both agents barely improve the TWAP benchmark out of sample, given that the \$  outperformance difference is reduced compared to the price of the asset (BTC had an average price close to 35k USD during the train/eval periods). This finding experimentally supports the results of \cite{almgren2000optimal,Bertsimas1998OptimalCO} which showed the TWAP to be the optimal execution algorithm under random walk dynamics. Both agents have trouble significantly generalizing the experience they have learned during training to a new market period because the environment presents a dynamic close to a random-walk and thus a low signal-to-noise ratio. Nonetheless, the FCN agent consistently achieves to outperform out of sample. It is however important to remember that our setup severely restricts the freedom of both agents. Allowing them to trade more freely, across different timestamps, and with more options for choosing prices, order types, and distributing volumes, could perhaps allow for a greater outperformance (likely with a bigger variance). Additionally, our limit order execution Algorithm \ref{LimitOrderExeAlgo} induces a delay in order execution, only executing volume if the market moves in our favor. Choosing a less restrictive simulation of order executions could also lead to easier dynamics for the agents to learn, but may not be as representative of a real market environment.\\
 
\begin{table}[]
    \centering
    \begin{tabular}{c c c c}
    \toprule
    & \multicolumn{3}{c}{\textbf{Mean $\pm$ std}} \\\cmidrule{2-4}  \textbf{Training Period} & 01/06/2021-05/06/2021 & 06/06/2021-10/06/2021 & 11/06/2021-16/06/2021  \\\cmidrule{2-4}  \textbf{Evaluation Period} & 06/06/2021-10/06/2021 & 11/06/2021-16/06/2021  & 16/06/2021-20/06/2021 \\\cmidrule{2-4} \textbf{Agent Name} & & & 
    \\\toprule
    \textbf{LSTM agent} & \textbf{0.0616 $\pm$ 0.48} & -0.0080 $\pm$ 0.33 & 0.0418 $\pm$ 0.34\\\midrule
    \textbf{FCN agent} & 0.0506 $\pm$ 0.46 & \textbf{0.0785 $\pm$ 0.39} & \textbf{0.1087 $\pm$ 0.40}
    \\\midrule
    \bottomrule\\
    \end{tabular}
    \caption{Out-Of-Sample performance of the iteratively trained models measured as the mean of the \$ difference between the RL achieved price (VWAP) vs that of the TWAP benchmark. Dates are in \textit{dd/mm/yyyy} format.}
    \label{table:Iterative Performance Table}
\end{table}

\subsection{Performance Decay Experiment}

In order to estimate the out of sample performance time decay of a model, we take the FCN agent of the previous experiment trained from 01/06/2021 to 05/06/2021 and evaluate it on the two other evaluation periods, from 11/06/2021 to 16/06/2021 and from 16/06/2021 to 21/06/2021. As in the previous experiment, we evaluate the agent for $10^4$ Episodes on each period, with the same execution parameters as during training. The results are summarized in Table \ref{table: Performance Decay Table}. We can observe how, unlike its iteratively trained counterparts, the agent trained only once does not improve its performance on the later periods but rather performs similarly, not significantly decaying.

\begin{table}[]
    \centering
    \begin{tabular}{c c c c}
    \toprule
     & \multicolumn{3}{c}{\textbf{Mean $\pm$ std}} \\\cmidrule{2-4}
     \textbf{Training Period} & & 01/06/2021-05/06/2021 &
     \\\cmidrule{2-4}  \textbf{Evaluation Period} & 06/06/2021-10/06/2021 & 11/06/2021-16/06/2021  & 16/06/2021-20/06/2021 \\\cmidrule{2-4} \textbf{Agent Name} & & & 
    \\\toprule
    \textbf{FCN agent} & 0.0506 $\pm$ 0.46 & 0.0586 $\pm$ 0.42 & 0.0566 $\pm$ 0.34 \\\midrule
    \bottomrule\\
    \end{tabular}
    \caption{Out-Of-Sample performance of a model trained on a single period measured via the mean of the \$ difference between the RL achieved price (VWAP) vs that of the TWAP benchmark. Dates are in \textit{dd/mm/yyyy} format.}
    \label{table: Performance Decay Table}
\end{table}

\subsection{Hardware Considerations}

For training and testing, we used a Nvidia DGX station equipped with a 40-thread CPU, 4x Tesla V100 32GB GPUs, and 256 GB of RAM. We observed that during the training of the agent, there is no significant speed-up when the agent's policy network is using the GPU. Thus, we used only the CPU for sampling the RL environment and also for training the agent. A couple of reasons could explain the lack of speed-up increase when trying to use the GPU. First, the policy network used by our RL agent is not big enough to take full advantage of the Nvidia Tesla V100. Second, there is an overhead data transfer between the sampling of the environment and the training of the policy. Because our environment was developed only to be used on the CPU (as it is usually the case in RL), there is a penalty when the information is transferred from the RAM (CPU) memory to the GPU memory and vice-versa. In addition, modern processors like ours are capable of doing efficient computation by leveraging the Advanced Vector Extensions 512 (Intel AVX-512) instructions when working with tensors.

\section{Conclusions and Outlook}

In this article, we have presented a modular framework for the implementation of different RL Optimal Execution setups. The framework can accommodate setups with different functionalities and simulation choices such as data sources, order execution simulation procedures, execution algorithms with multiple types of actions, arbitrary order submission schedules, and reward calculation functions among others. We have presented each module together with their intended functionalities and the methods they are required to incorporate. We have stated the dependencies between modules and presented a concrete implementation of an example setup that aims to improve a TWAP benchmark through RL. The example setup trains a narrow class of execution algorithms that follow a TWAP order submission schedule. We have shown how to implement this through specific modules and discussed the explicit and implicit effects that each module induces on the simulation. In our experiments, we have trained and evaluated two different agents, over multiple market periods, shown the out of sample outperformance of both models to be small due to the restrictive nature of the environment and finally shown the performance of the best performing model not to significantly decay over time.

We have made our implementation as well as the code necessary to replicate the experiments publicly available. In future work, we plan on leveraging the framework to implement more complex simulation setups that tackle the deficiencies discussed in Section \ref{Section3 Setup Example}. Since ranking setups by their realism is not free of challenges, a good approach to selecting agents to carry out live executions will be to choose those that can consistently outperform benchmarks over the biggest number of setups.

\bibliographystyle{plain} 
\bibliography{references}

\end{document}